# Navigating the Docker Ecosystem: A Comprehensive Taxonomy and Survey

Prathamesh Muzumdar [a*], Amol Bhosale [b],
Ganga Prasad Basyal [c] and George Kurian [d]

[a] *The University of Texas at Arlington, USA.*
[b] *Capgemini, Malaysia.*
[c] *Black Hills State University, USA.*
[d] *Eastern New Mexico University, USA.*

*Authors' contributions*

*This work was carried out in collaboration among all authors. All authors read and approved the final manuscript.*

Review Article

## ABSTRACT

The cloud computing landscape is rapidly expanding and growing in complexity. It has witnessed the emergence of Cloud Computing as a widely adopted model for efficiently processing large volumes of data by harnessing clusters of commodity computers. This evolution enables the handling of massive data through on-demand services, relying on numerous microservices with diverse dependencies. The technology of containers ensures secure storage, allowing for large-scale data processing with high scalability and portability. Container technology, particularly exemplified by Docker in the last decade, plays a pivotal role in this scenario. It empowers microservices to process data swiftly, enabling developers to dynamically scale these services in real-time. This paper initiates by establishing a comprehensive taxonomy for delineating container architecture. Focusing specifically on Docker containers, we scrutinize various existing container-related literature. Through this taxonomy and survey, we not only discern similarities and disparities in the architectural approaches of Docker container technology but also pinpoint areas necessitating further research.

_________________________________________________________________________________

*Corresponding author: E-mail: prathameshmuzumdar85@gmail.com;*

## 1. INTRODUCTION

Has Container technology merely generated buzzwords and hype, or does it hold substantial value? Beyond the surface excitement, Container technology emerges as a profoundly transformative force, gaining traction due to its scalability and portability [1]. It inherits established technology while introducing innovative concepts to construct an efficient system, surpassing the conventional mainframe model. Container technology embodies the next evolutionary phase in distributed computing [3]. This model aims to optimize the utilization of distributed resources, amalgamating them to achieve heightened throughput and tackle extensive computational challenges at a micro level, exemplified by microservices [11]. Notably, container technology isn't an entirely novel notion in the realm of developing and operating large-scale web applications within a service-oriented architecture (SOA) [6]. It facilitates the cost-effective development of scalable web portals on robust, fault-tolerant infrastructures.

Containerization, a lightweight virtualization technology, has revolutionized the management of cloud applications [27]. It provides a means to encapsulate applications and their dependencies in isolated environments, ensuring seamless deployment across various platforms. As this technology gained prominence, the challenge of efficiently orchestrating the creation and deployment of containers, both individually and in clusters, emerged as a pivotal concern [18]. The adoption of containers, exemplified by Docker, has propelled the development and deployment of applications, offering benefits such as portability, resource efficiency, and rapid scaling [22]. However, orchestrating these containers in a coordinated manner, especially in complex, multi-container environments, demands sophisticated solutions.

Container orchestration platforms like Kubernetes have risen to prominence to address this very challenge [12]. They provide robust tools for automating the deployment, scaling, and management of containerized applications. Kubernetes, in particular, has become the de facto standard for container orchestration, offering features like service discovery, load balancing, and automatic scaling based on resource usage [28]. In summary, containerization technology has ushered in a new era of application deployment, but effective orchestration of containers, whether individually or in clusters, has become a critical concern in modern IT operations [16,23]. This has led to the rise of advanced orchestration platforms like Kubernetes, which play a central role in managing containerized applications at scale.

Our research has culminated in a comprehensive repository of contemporary research methodologies, techniques, best practices, and real-world experiences employed in the realm of container technology architecture, with a specific emphasis on the development and management of containerized applications and microservices [36]. Through our investigation, it became evident that container technology research is currently undergoing a phase of foundational development, necessitating further experimentation and empirical assessment of its advantages [35]. Additionally, our study led to the formulation of an all-encompassing classification system tailored specifically for container technology, particularly within the context of clustered cloud architectures [25].

The findings from our mapping study reveal a growing interest and adoption of container-based technologies, such as Docker, which serve as lightweight virtualization solutions at the infrastructure-as-a-Service (IaaS) level, and as application management tools at the Platform-as-a-Service (PaaS) level. It is evident that containers have a positive impact on both the development and deployment phases [21]. For example, cloud architecture is shifting towards DevOps-driven methodologies, supporting a seamless cycle of continuous development and deployment that leverages cloud-native solutions based on container technology and their orchestration.

The results demonstrate that containers play a pivotal role in facilitating continuous development in the cloud, utilizing cloud-native platform services for development and deployment. However, they do require sophisticated orchestration support, which can be provided by platforms like Docker Swarm or Kubernetes. Consequently, container-based orchestration techniques emerge as crucial mechanisms for coordinating computation in cloud-based, clustered environments [24]. Our study highlights

Docker as a superior container technology when compared to alternatives such as LXC (Linux containers), Windows Hyper-V, Podman, runC, and containerd.

## 2. BACKGROUND

Virtualization of resources involves using an additional layer of software above the host operating system to manage multiple resources. These virtual machines (VMs) function as distinct execution environments. Various approaches are employed for virtualization, with one popular method being hypervisor-based virtualization [32]. Notable solutions in this category include KVM and VMware. To implement this technology, a virtual machine monitor is necessary above the underlying physical system, and each VM provides support for isolated guest operating systems [10]. It's conceivable for a single host operating system to accommodate numerous guest operating systems within this virtualization framework.

On the other hand, container-based virtualization represents a different approach. Here, hardware resources are partitioned to create multiple instances with secure isolation properties [8]. The key distinction lies in how guest processes interact with the technology. With container-based systems, processes access abstractions directly through the virtualization layer at the operating system (OS) level. In contrast, hypervisor-based approaches typically have one VM per guest OS. Container-based solutions often share a single OS kernel among virtual instances, leading to an assumption of weaker security compared to hypervisors [31]. From a user's perspective, containers function as self-contained operating systems, seemingly capable of independent operation from both hardware and software.

Containerization enables a streamlined form of virtualization by crafting specialized containers from distinct images (typically sourced from an image repository) as self-contained application bundles. These containers utilize fewer resources and time. Moreover, they facilitate a higher level of compatibility in packaging applications, crucial for creating portable and interoperable software solutions in cloud environments. The foundation of containerization lies in the ability to efficiently build, assess, and launch applications across numerous servers, as well as establishing connections between these containers [38]. As a result, containers effectively tackle concerns at the cloud Platform as a Service (PaaS) level.

## 3. LITERATURE SURVEY

Container technology has emerged as a transformative force in modern computing, garnering attention for its scalability and portability [1]. Unlike traditional approaches, it introduces innovative concepts that construct efficient systems, surpassing the limitations of conventional mainframe models [1]. This review explores the evolution and impact of container technology, shedding light on its role in distributed computing and its applications in managing cloud resources.

Container technology represents the next evolutionary phase in distributed computing [3], aiming to optimize the utilization of distributed resources for heightened throughput, particularly in the context of microservices [11]. While not entirely novel, container technology has proven effective in developing and operating large-scale web applications within service-oriented architectures (SOA) [6]. It facilitates cost-effective development on robust, fault-tolerant infrastructures, positioning itself as a key player in modern software architecture.

Containerization, a lightweight virtualization technology, has revolutionized the management of cloud applications [27]. It provides a means to encapsulate applications and their dependencies in isolated environments, ensuring seamless deployment across various platforms. However, the efficient orchestration of container creation and deployment, especially in complex, multi-container environments, has emerged as a pivotal concern [18]. This has led to the rise of container orchestration platforms, with Kubernetes standing out as the de facto standard, addressing challenges in automating deployment, scaling, and management of containerized applications [12].

Research in container technology has reached a phase of foundational development, prompting the need for further experimentation and empirical assessment of its advantages [35]. Comprehensive studies have led to the formulation of a classification system tailored for container technology within clustered cloud architectures [25]. These developments underscore the dynamic nature of container technology research, requiring ongoing exploration and validation of its potential.

Mapping studies reveal a growing interest and adoption of container-based technologies, such as Docker, serving as lightweight virtualization solutions at the Infrastructure as a Service (IaaS) level and as application management tools at the Platform as a Service (PaaS) level [21]. Containers positively impact both development and deployment phases, aligning with the shift towards DevOps-driven methodologies in cloud architecture [21]. Continuous development and deployment cycles leverage cloud-native solutions based on container technology and their orchestration, emphasizing their role in shaping modern IT operations.

Containers play a pivotal role in facilitating continuous development in the cloud, utilizing cloud-native platform services for development and deployment [24]. While they offer benefits in terms of resource efficiency and rapid scaling, sophisticated orchestration support is essential. Platforms like Docker Swarm or Kubernetes emerge as crucial mechanisms for coordinating computation in cloud-based, clustered environments [24]. Among various container technologies, Docker is highlighted as a superior choice when compared to alternatives [24].

Transitioning to the broader context of virtualization, the review explores the conventional approach of using an additional layer of software, such as hypervisors like KVM and VMware, above the host operating system to manage multiple resources [32]. These hypervisor-based virtualization solutions enable the creation of distinct execution environments, each supporting isolated guest operating systems. The ability of a single host operating system to accommodate numerous guest operating systems within this virtualization framework demonstrates its flexibility and scalability.

In contrast, container-based virtualization represents a different approach, where hardware resources are partitioned to create multiple instances with secure isolation properties [8]. The key distinction lies in how guest processes interact with the technology, with containers providing direct access through the virtualization layer at the operating system level. This differs from hypervisor-based approaches, where each VM typically corresponds to one guest OS. Container-based solutions often share a single OS kernel among virtual instances, leading to an assumption of weaker security compared to hypervisors [31].

Containerization enables a streamlined form of virtualization by crafting specialized containers from distinct images, serving as self-contained application bundles [38]. These containers offer advantages in terms of efficiency, utilizing fewer resources and time. Their compatibility in packaging applications is crucial for creating portable and interoperable software solutions in cloud environments. The ability to efficiently build, assess, and launch applications across numerous servers, while establishing connections between containers, positions containers as effective solutions at the cloud Platform as a Service (PaaS) level [38].

In conclusion, container technology has transcended buzzwords and hype, demonstrating substantial value in modern computing. Its transformative force, scalability, and portability have propelled it to the forefront of distributed computing and cloud management. While challenges exist, the evolution of containerization, coupled with advanced orchestration platforms like Kubernetes, signifies a new era in application deployment and management. Additionally, the comparison with hypervisor-based virtualization highlights the unique advantages and considerations associated with each paradigm, offering insights into the diverse landscape of virtualization technologies. The ongoing research and development in container technology further emphasize its dynamic nature, calling for continuous exploration and validation of its potential in reshaping the IT landscape.

## 4. TAXONOMY OF DOCKER CONTAINER

Numerous taxonomies exist for cloud computing, but few are dedicated to container systems like Docker. Our taxonomy serves as a valuable resource for academia, developers, and researchers, offering insights into container architecture and management [15]. We developed this taxonomy by considering both vendors within the Cloud Container landscape and enterprise IT, the end-users of cloud services and software. In the following sections, we present and elaborate on this taxonomy, providing a comprehensive understanding of the subject.

## 4.1 Docker Architecture

Fig. 1 depicts the all-encompassing structure of Docker container architecture. It encompasses the design of software applications that utilize internet-accessible, on-demand services. Docker Architectures rely on an infrastructure that is activated only when required, pulling the essential services (microservices) as needed to execute a specific task, and subsequently releasing any surplus resources, often discarding them upon task completion [14].

The architecture of a container cluster comprises one or more master nodes and one or more worker nodes. The worker nodes function as distinct clusters that deliver microservices organized according to grouped business logic [20]. API services facilitate communication between containers, while configuration services, specifically Docker configs, serve as repositories for non-sensitive information, like configuration files, external to a service's image or active containers [34]. This approach enables you to maintain your images in a more universal manner, eliminating the necessity to bind-mount configuration files into the containers or rely on environment variables [43].

Numerous internal services facilitate communication, alongside a variety of external services directing business inquiries to a specific service within a container or a group of containers grouped in a pod within a particular cluster [7]. An external client engages with diverse services through a master node, which interacts with the internal cluster loop, redirecting requests to specific services housed in a pod on a worker node. A load balancer aids in pinpointing an available cluster prepared to deliver the service. The master node encompasses distinct configurations, including an API server, Scheduler, controller manager, and ETCD.

## 4.2 Docker Client Server Architecture

### 4.2.1 Docker daemon

Docker Daemon, often referred to simply as "Dockerd," is a crucial component at the heart of Docker container technology. It serves as the background service responsible for managing and executing Docker containers on a host system. Docker Daemon listens for Docker API requests and translates them into actions, facilitating container creation, management, and networking [30]. This lightweight, long-running process plays a pivotal role in container orchestration, resource allocation, and security enforcement. It ensures containers run efficiently, isolating them from the host system while offering a unified interface for developers and operators. Docker Daemon is an integral part of the Docker ecosystem, enabling the agility, scalability, and portability that containerization is renowned for.

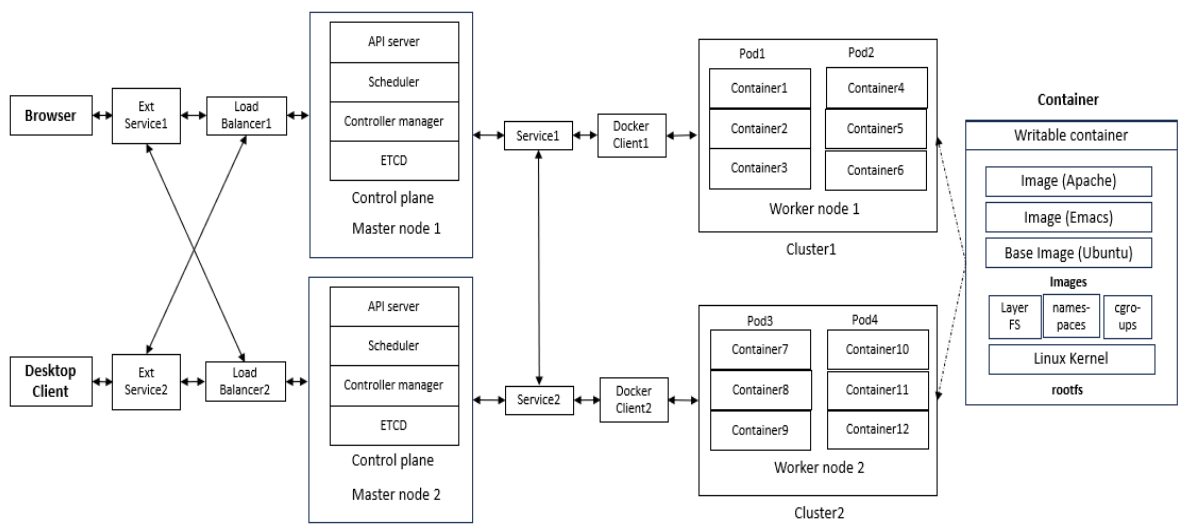

**Fig. 1. Docker Architecture [43]**

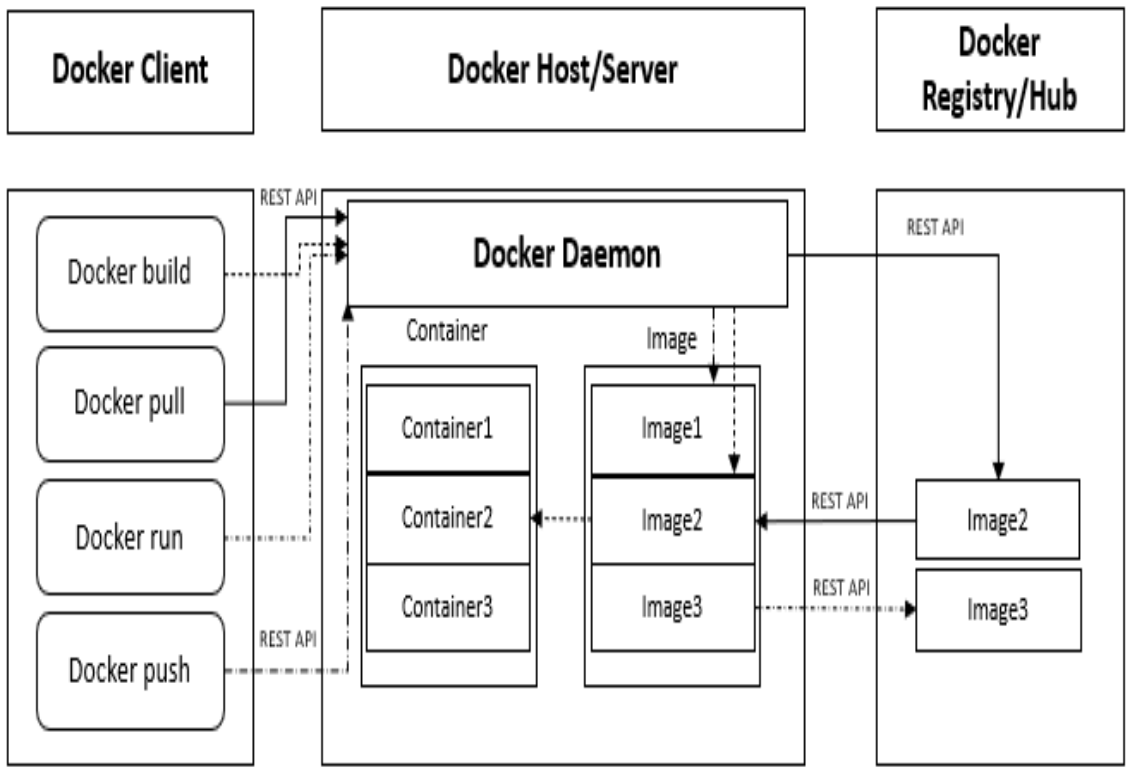

**Fig. 2. Docker Client Server Architecture [43]**

**4.2.2 Docker client**

The Docker Client is a command-line interface (CLI) that allows users to interact with Docker Daemon, facilitating container operations. It sends commands to the daemon via the Docker API, enabling tasks like building, running, and managing containers. The client is pivotal for developers and administrators using Docker technology.

**4.2.3 Docker build**

Docker Build is a pivotal command in Docker's toolkit, facilitating the creation of custom container images. It employs a Dockerfile—a declarative script defining the image's configuration and dependencies. When executed, Docker Build reads this file, then compiles and packages the application along with its environment. Each step is cached, enhancing efficiency during subsequent builds. This process ensures reproducibility and portability across different environments. Docker Build is fundamental for DevOps workflows, enabling seamless integration with version control systems. It empowers developers to encapsulate applications and their dependencies, ensuring consistent deployment in diverse computing environments.

**4.2.4 Docker pull**

Docker Pull is a command that allows users to fetch container images from a specified registry, typically Docker Hub or a private repository. When executed, it contacts the Docker Daemon, which then downloads the requested image and its layers, if they aren't already present on the local system [37]. This process ensures that the required image is readily available for running containers. Docker Pull is a vital aspect of containerization, enabling rapid deployment by retrieving pre-configured images. It streamlines software distribution, making it a cornerstone in modern development and deployment pipelines, fostering efficiency and consistency across diverse computing environments.

**4.2.5 docker run**

Docker Run is a fundamental command in Docker, enabling the execution of containerized applications. It instructs Docker to create and initiate a new container based on a specified image. Users can customize container behavior through various options, such as network settings, environment variables, and resource constraints. Once initiated, the container runs as an isolated instance, leveraging the host system's resources while maintaining its own file system and networking. Docker Run is instrumental in achieving consistency across development, testing, and production environments, providing a seamless deployment process. It's a linchpin in modern software development, promoting portability, scalability, and efficient resource utilization.

### 4.2.6 Docker push

Docker Push is a pivotal command enabling users to upload their custom-built container images to a container registry, like Docker Hub or a private repository. This process involves tagging the image appropriately, specifying the destination registry, and authenticating the user's credentials [4]. Once executed, Docker Push transfers the image and its layers to the specified repository, making it accessible to others in the development or operations team. This capability streamlines collaboration and deployment workflows, ensuring that consistent and tested images are readily available for deployment across diverse environments [33]. Docker Push is a linchpin in modern DevOps practices, fostering efficiency, scalability, and reproducibility.

## 4.3 Docker Container

Docker containers revolutionize software deployment by encapsulating applications and their dependencies in isolated units. These lightweight, portable environments ensure consistency across different computing environments, from development to production. Unlike virtual machines, Docker shares the host OS kernel, consuming fewer system resources and enabling rapid startup times. This efficiency allows for efficient utilization of server resources, enabling more containers to run on a single machine. Docker's image-based approach facilitates version control and easy replication, ensuring reliable and reproducible deployments. With a vast ecosystem and support for orchestration tools like Kubernetes, Docker has become a cornerstone of modern DevOps practices, fostering scalability, agility, and seamless deployment pipelines.

Docker containers and images are interdependent components of the Docker platform. An image serves as a lightweight, standalone, and immutable template containing an application and its dependencies. It's a snapshot of a specific environment, ensuring consistency across various stages of development and deployment [40]. When a container is instantiated from an image, it becomes a runnable instance of that environment. Containers are dynamic, allowing applications to execute, interact, and communicate with other services. They also introduce a writable layer, enabling changes to be saved during runtime. Images serve as blueprints, while containers bring them to life, collectively enabling the agility, portability, and reproducibility that Docker is celebrated for.

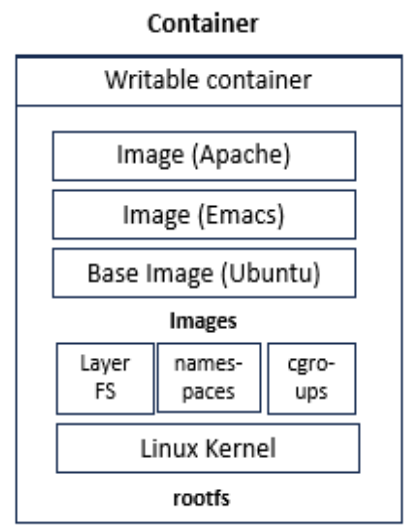

**Fig. 3. Docker Container [43]**

## 4.4 Docker Image

A Docker image is a fundamental component of containerization, encapsulating an application and its dependencies into a portable, immutable package. These images serve as lightweight templates, offering consistency across various environments, from development to production. Docker images are based on a layered file system, with each layer representing a specific instruction in the image's Dockerfile. This layered approach allows for efficient storage and sharing of common components. Images are versioned, making it easy to track changes and ensure reproducibility in deployments [26]. They play a pivotal role in modern DevOps workflows, facilitating seamless application packaging, distribution, and scalability while promoting efficient resource utilization.

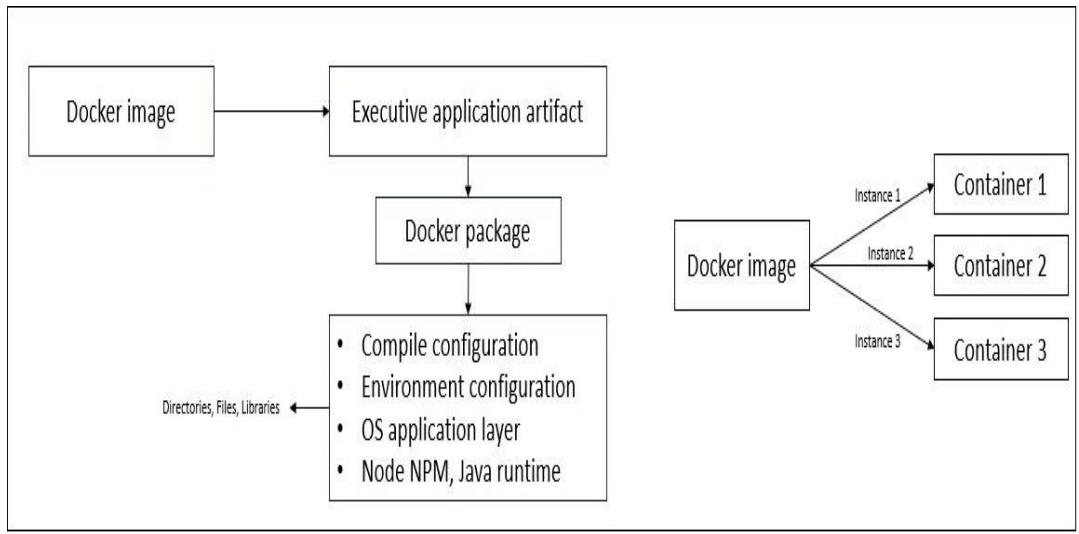

**Fig. 4. Docker Image [43]**

### 4.5 Docker Package

Docker packages revolutionize software deployment through containerization. A Docker package comprises an application, its dependencies, libraries, and runtime environment, encapsulated into a self-contained unit. This package, known as a Docker image, is lightweight and portable, ensuring consistent performance across different computing environments. Docker's layered file system allows for efficient storage and sharing of components, optimizing resource usage.

Images can be versioned, enabling precise control over deployments. With Docker, packages become highly reproducible, streamlining development to production workflows [41]. This transformative technology has become a cornerstone of modern DevOps, driving efficiency, scalability, and agility in software development and deployment processes.

### 4.6 Docker Registry (Docker Hub)

A Docker registry is a vital component of the Docker ecosystem, acting as a centralized repository for Docker images. It serves as a storage and distribution hub for containerized applications, allowing users to securely share, manage, and retrieve Docker images. Registries play a crucial role in facilitating collaboration among developers and teams, ensuring seamless deployment workflows [42]. Docker Hub is one of the most widely used public Docker registries. It provides a cloud-based platform for sharing and accessing Docker images. Docker Hub hosts a vast library of pre-built images for various applications, operating systems, and development stacks [29]. This extensive collection accelerates the development process by eliminating the need to build images from scratch. Developers can simply pull ready-to-use images from Docker Hub, saving time and effort.

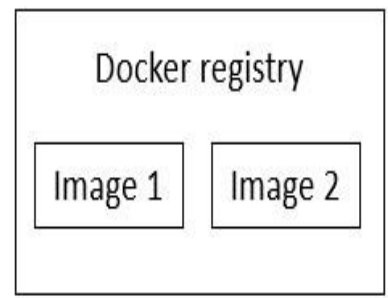

**Fig. 5. Docker Registry [43]**

Additionally, Docker Hub supports version control, enabling users to manage different iterations of their images. It also offers tools for automated builds, allowing for continuous integration and delivery workflows. Docker Hub's user-friendly interface and robust search capabilities simplify the process of discovering and incorporating images into projects. While Docker Hub is immensely popular, organizations often deploy private Docker registries for enhanced security and control over their image repositories [5]. These private registries, like Docker Trusted Registry (DTR), enable teams to securely store proprietary images and integrate them seamlessly into their development pipelines.

### 4.7 Docker Repository

A Docker repository is a storage location for Docker images. It serves as a centralized hub where containerized applications, along with their dependencies and configurations, are stored. Repositories allow for efficient management, sharing, and retrieval of Docker images. They come in two types: public and private [39]. Public repositories, like Docker Hub, are open to the community, providing a vast collection of pre-built images for various applications. Private repositories, on the other hand, are typically used by organizations to securely store and manage proprietary images. They offer control over access and permissions, ensuring sensitive applications remain confidential within a team or company.

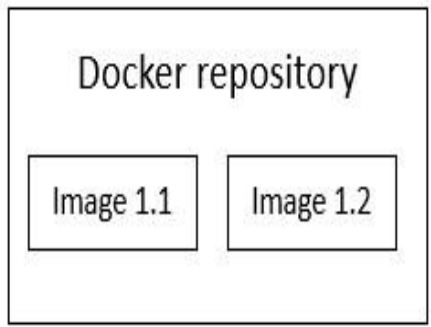

**Fig. 6. Docker Repository [43]**

### 4.8 Port Binding

Port binding is a crucial concept in networking, particularly in the context of containerization technologies like Docker. It involves connecting specific network ports of a container to the host system or other containers. This enables external services to communicate with applications running inside the container. By binding container ports to the host, developers can ensure seamless interaction between the containerized application and external resources. Port binding plays a pivotal role in orchestrating microservices architectures, allowing different components to communicate effectively [9]. It also enhances security by controlling which ports are exposed externally, safeguarding sensitive services from unauthorized access. This practice is integral to building robust, interconnected systems in modern software development.

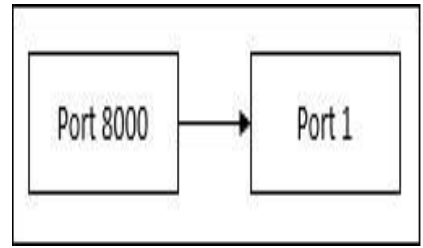

**Fig. 7. Port Binding [43]**

### 4.9 Docker File

A Dockerfile is a crucial component in the Docker ecosystem, serving as a blueprint for creating Docker images. It's a plain-text configuration file that contains a series of instructions, specifying how to assemble an environment within a Docker container. These instructions encompass actions like specifying a base image, adding files, running commands, and configuring settings. The Dockerfile's layered approach is fundamental to Docker's efficiency. Each instruction creates a new layer in the image, optimizing resource usage and enabling faster builds. This also facilitates image versioning, allowing developers to track changes and roll back to previous states if needed. Developers use Dockerfiles to automate the process of setting up environments, ensuring consistency across development, testing, and production stages [17]. This automation minimizes the "it works on my machine" problem, fostering reproducibility in deployments.

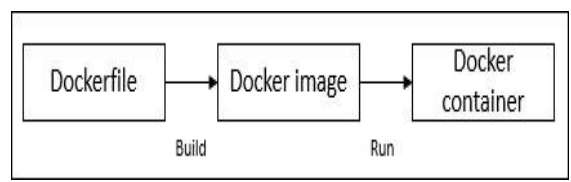

**Fig. 8. Docker File [43]**

Moreover, Dockerfiles promote best practices like keeping images lightweight by minimizing unnecessary components, enhancing security by only including essential dependencies, and improving efficiency by leveraging cache for repetitive tasks. A well-crafted Dockerfile is a cornerstone of efficient containerized workflows [13]. It empowers developers to encapsulate applications and their dependencies, making them easily portable and deployable across different computing environments. This enables rapid, reliable, and scalable application development and deployment, making Dockerfiles an indispensable tool in modern DevOps practices.

## 4.10 Deployment Server

A Docker deployment server is a critical component in the lifecycle of containerized applications, responsible for managing the distribution, scaling, and orchestration of Docker containers in a production environment. It plays a pivotal role in ensuring that applications run smoothly, efficiently, and reliably at scale. One of the most popular tools for Docker deployment is Kubernetes, an open-source container orchestration platform. Kubernetes abstracts the underlying infrastructure and provides a declarative way to define how containerized applications should be deployed, scaled, and managed [2]. It automates tasks like load balancing, scaling, rolling updates, and self-healing, making it ideal for large-scale and complex container deployments.

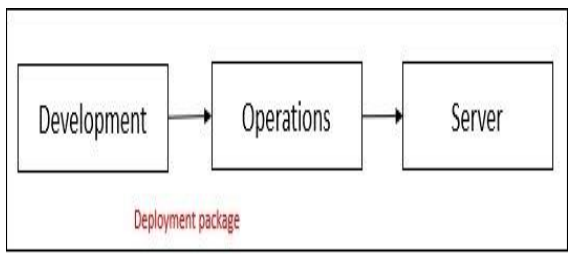

**Fig. 9. Deployment Server [43]**

Another commonly used tool is Docker Swarm, Docker's built-in orchestration solution. Docker Swarm simplifies the deployment and scaling of Docker containers and is well-suited for smaller, less complex applications. In addition to these tools, various cloud providers offer managed container orchestration services, such as Amazon Elastic Kubernetes Service (EKS), Google Kubernetes Engine (GKE), and Azure Kubernetes Service (AKS), which make it easier to deploy and manage containerized applications in the cloud. A Docker deployment server, whether based on Kubernetes, Docker Swarm, or a cloud service, is essential for achieving high availability, fault tolerance, and efficient resource utilization in containerized environments [19]. It allows organizations to deploy and manage applications with confidence, knowing that their containerized workloads are being handled effectively.

## 4.11 Docker Bind Mount

Docker bind mount is a technique that allows a file or directory on the host system to be mounted into a container. This establishes a direct link between the container and the host's file system, enabling real-time synchronization of files and data. Bind mounts are versatile, supporting read and write operations, making them invaluable for development workflows, debugging, and sharing data between a container and its host [16]128]. They offer flexibility and simplicity, allowing containers to access and modify host files without the need to create custom images [22]. Docker bind mounts are a powerful tool for seamless integration between containerized applications and the host environment.

## 4.12 Docker Stack

A Docker stack is a collection of services, often comprising multiple containers, that work together to deliver a cohesive application [18]. It's defined using a Compose file, which outlines the configuration for each service, including container images, networking, and dependencies. Docker Swarm, Docker's native orchestration tool, manages these stacks, ensuring high availability, fault tolerance, and scalability. Docker stacks facilitate the deployment of complex, multi-container applications in a simplified manner [38]. They offer benefits like service discovery, allowing containers within a stack to communicate seamlessly, and load balancing, distributing traffic evenly across the service instances.

Furthermore, Docker stacks enable automated rollouts and rollbacks, making it easy to update applications while minimizing downtime. They also support health checks, ensuring that only healthy containers receive traffic. In addition to these features, Docker stacks are compatible with overlay networks, enabling communication across multiple hosts in a Swarm cluster. This enables the creation of resilient, distributed applications [23]. Overall, Docker stacks are instrumental in orchestrating the deployment of multi-container applications, providing an efficient and scalable framework for modern software architectures. They streamline the management of complex applications, making them an essential tool for DevOps teams and organizations embracing containerization.

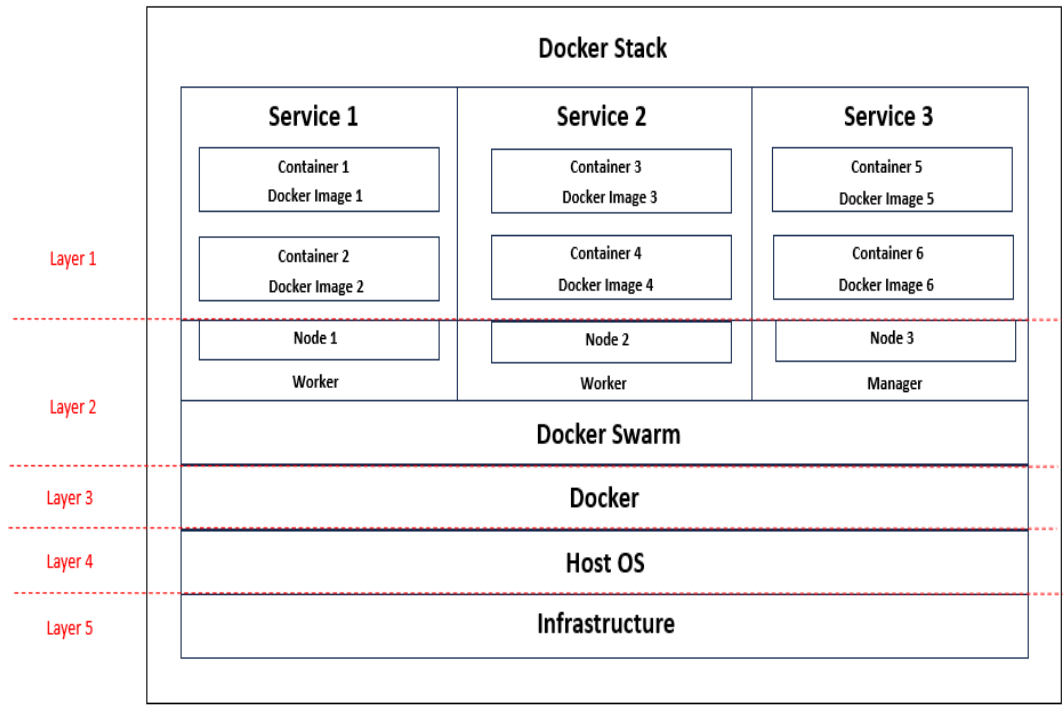

**Fig. 10. Docker Stack [43]**

### 4.13 Docker Deployment and Deployment Process

*Docker Deployment and Deployment Process:*

Docker has transformed the way applications are deployed, providing a consistent and portable environment throughout the entire software development lifecycle [16]. Docker deployment involves the process of taking Docker containers, typically packaged as Docker images, and running them in various environments, from development to production. Here's an overview of the Docker deployment process:

*1. Building Docker Images*

Developers create Docker images by writing a Dockerfile that defines the application, its dependencies, and configuration [10]. These images are built using the `docker build` command, which generates a snapshot of the application and its environment.

*2. Registry and Image Storage*

Docker images are stored in Docker registries like Docker Hub, private registries, or container image repositories [11]. Registries serve as centralized repositories where images can be pushed, pulled, and versioned.

*3. Container Orchestration*

In production, Docker containers are often managed and orchestrated using tools like Docker Swarm or Kubernetes [22]. Orchestration platforms automate the deployment, scaling, load balancing, and monitoring of containers, ensuring high availability.

*4. Deployment Configuration*

Deployment configurations are defined using YAML files (Docker Compose for development or Kubernetes YAML files for production) [32]. These files specify how containers should be run, networked, and linked together.

*5. Deployment Stacks*

Applications may consist of multiple services or microservices defined in a Docker stack [24]. Docker stacks simplify the management of multi-container applications, ensuring they work cohesively.

*6. Rollouts and Updates*

Docker allows for seamless application updates using strategies like rolling updates [32]. Containers can be updated with minimal downtime, and the process can be automatically reversed in case of issues.

*7. Service Discovery*

Docker provides service discovery mechanisms to enable containers to find and communicate with each other, essential for microservices architectures [12].

*8. Health Checks and Scaling*

Docker containers can be configured with health checks to ensure they are responsive and healthy [7].Orchestration platforms automaticallyscale containers up or down based on demand.

*9. Monitoring and Logging*

Docker deployments benefit from robust monitoring and logging solutions to track container performance, troubleshoot issues, and analyze application behavior [6].

*10. Security and Access Control*

Security measures like image scanning, container runtime protection, and access controls are crucial to secure Docker deployments [15]. Docker deployment simplifies application management, accelerates development, and promotes consistency across various environments. It's a core technology in modern DevOps practices, enabling efficient, scalable, and reliable software delivery.

## 5. SURVEY

While previous research has explored Docker container systems from industrial and informational viewpoints, a thorough technical investigation is notably absent. Our study delves into Docker container systems, comparing them with various other container systems. We employ diverse classification criteria in this section to facilitate a comprehensive technical comparative analysis. Table 1 presents the outcomes of this technical comparative study, showcasing the distinctions between various infrastructure and solution providers across different containers.

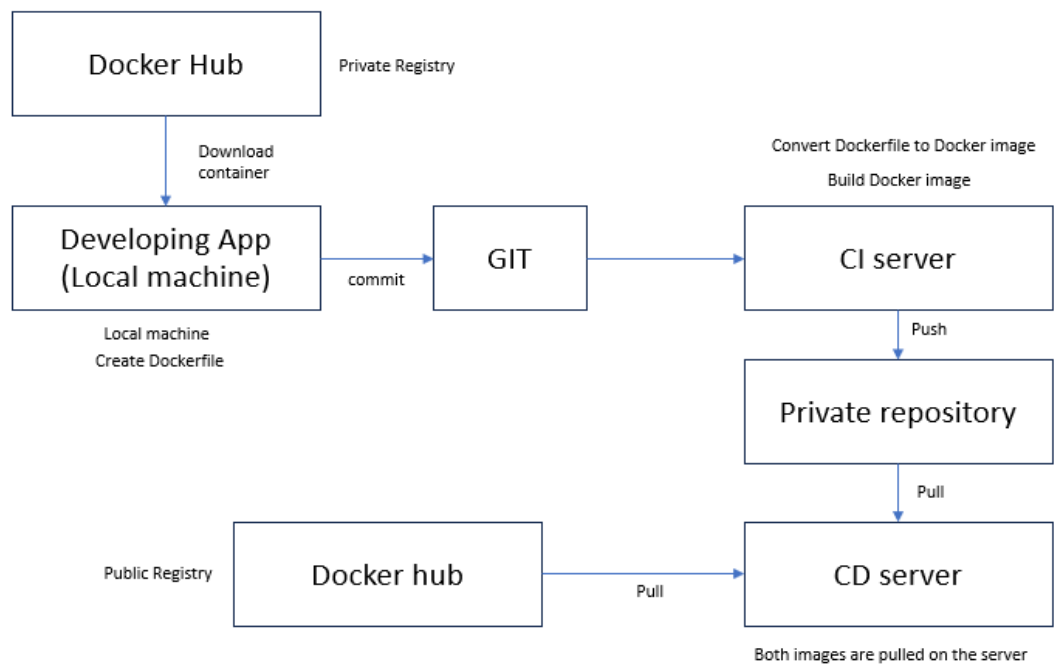

**Fig. 11. Docker Deployment Process [43]**

**Table 1. Comparative Table [7, 9. 21, 22, 24, 34, 43]**

| Criteria | Docker | LXC | Windows Hyper-V | Podman | runC | containerd |
|---|---|---|---|---|---|---|
| Design focus | Simplicity and user-friendly | Container Isolation and Efficiency | Isolation and virtualization of Windows environments | Lightweight and Daemonless | Container Runtime and Spec Compliance | Container Runtime and Spec Compliance |
| Microservices management system | Layered approach for deploying and scaling applications | N/A (LXC is more focused on low-level containerization) | N/A (Hyper-V is more focused on traditional virtualization) | N/A (Podman is more focused on lightweight container management) | N/A (runC is a low-level runtime) | Container Runtime Spec compliance |
| Portability | Consistent runtime environment on any platform | Limited to Linux-based systems | Limited to Windows-based systems | Consistent runtime environment on any platform | Container Runtime Spec compliance | Container Runtime Spec compliance |

**Chart 1. TAXONOMY**

| No. | Terms | Definitions |
|---|---|---|
| 1 | Docker container | Lightweight, portable, and self-sufficient unit that encapsulates an application and its dependencies |
| 2 | Docker command | Instructions used in the Docker platform's command-line interface (CLI) to interact with Docker and perform tasks |
| 3 | Docker image | Images serve as the blueprint for creating Docker containers. Built from a base image or another existing image through a series of defined instructions in a Dockerfile, and can be stored in registries, allowing for easy sharing and distribution |
| 4 | Docker package | Synonym for Docker container |
| 5 | Docker registry (Docker hub) | Storage platform for Docker images, facilitating their distribution, sharing, and deployment across different environments |
| 6 | Docker repository | Collection of related Docker images, often versioned and organized for efficient storage and retrieval |
| 7 | Docker file | Plain-text configuration file that contains instructions for building a Docker image, specifying the base image, dependencies, and setup steps for a containerized application |
| 8 | Docker bind mount | Allows a file or directory on the host system to be mounted directly into a container, enabling real-time synchronization and data sharing between the host and the container |
| 9 | Docker stack | A group of services defined in a Docker Compose file that work together to form a multi-container application, allowing for simplified deployment and management |
| 10 | Docker compose | A tool that allows users to define and manage multi-container Docker applications using a simple YAML configuration file, enabling easy orchestration of complex deployments |
| 11 | Docker logs | A real-time stream of output generated by a running container, offering valuable insights into its behavior and troubleshooting capabilities |
| 12 | Docker exec | It allows users to run commands inside a running container, providing a means for direct interaction and troubleshooting without the need to create a new instance |
| 13 | Docker stats | It provides real-time resource usage statistics for running containers, offering insights into CPU, memory, and network performance at a glance |
| 14 | Docker network | It is a virtual environment that allows containers to communicate with each other, enabling seamless interaction between services in a distributed application |
| 15 | Bridge | It is a default network mode that allows containers on the same host |

| No. | Terms | Definitions |
|---|---|---|
| | | to communicate with each other via a virtual network bridge, providing isolation and secure connectivity |
| 16 | Host | It refers to the system where Docker is installed and running, managing containers and providing the necessary resources for their execution |
| 17 | Overlay | A network that allows containers across multiple hosts to communicate seamlessly, enabling distributed applications to function as a unified system |
| 18 | None | The "none" network mode in Docker isolates a container, preventing it from having any network access |
| 19 | Macvlan | It allows a container to have its own MAC address and appear as a physical device on the network, enabling direct communication with external systems |
| 20 | Docker inspect | It is a command-line tool that provides detailed information about Docker objects like containers, networks, volumes, or images, aiding in troubleshooting and configuration analysis |
| 21 | Docker image history | It displays the layers and changes within an image, providing insights into its creation and evolution |
| 22 | Docker image prune | It is a command used to remove unused or dangling images, reclaiming storage space |
| 23 | Docker container prune | It is a command used to remove stopped containers, freeing up system resources and storage space |
| 24 | Docker compose up | It starts and orchestrates the services defined in a Docker Compose file, creating a multi-container environment |
| 25 | Docker compose down | It stops and removes the services defined in a Docker Compose file, terminating the multi-container environment |
| 26 | Docker compose ps | It lists the status of services defined in a Docker Compose file, indicating whether they are running or stopped |
| 27 | Docker compose logs | It displays the combined output of services defined in a Docker Compose file, aiding in monitoring and troubleshooting |
| 28 | Docker storage | It manages the persistent data associated with containers, providing options for storage drivers, volumes, and storage management strategies |
| 29 | Data volumes | It provides a way to persistently store and share data between containers and the host system |
| 30 | Volume container | It is a dedicated container created solely for the purpose of managing shared data volumes, providing a centralized storage solution for other containers |
| 31 | Directory mounts | It allows specific host directories to be linked with containers, enabling real-time synchronization of data |
| 32 | Storage plugins | It extends the platform's capabilities by allowing different storage drivers to manage container data, supporting various backend technologies |
| 33 | Advanced Multi-Layered Unification File system (AuFS) | It is a union filesystem used by Docker to layer images, providing a lightweight and efficient way to manage container file systems |
| 34 | Compile configurations | It involves specifying instructions in a Dockerfile to build custom images with specific settings, dependencies, and behaviors |
| 35 | Environment configurations | It refers to variables and settings that can be passed to containers at runtime, influencing their behavior and functionality |
| 36 | Artifact | An artifact in software development is a result of a build process, which may include compiled code, libraries, or other files, ready for deployment |
| 37 | Port binding | It involves mapping a container's internal port to a specific port on the host system, enabling external access to the containerized service |
| 38 | CI server | Continuous Integration server, automates the process of code integration, testing, and deployment, ensuring a streamlined and reliable software development workflow |
| 39 | Deployment server | It is a dedicated platform that manages the distribution and installation of software applications to target environments, ensuring consistent |

| No. | Terms | Definitions |
|-----|-------|-------------|
|     |       | and reliable deployments |
| 40  | Kernel namespaces | It provides process isolation and resource control, allowing multiple instances of the same resource to coexist independently in a Linux environment |
| 41  | cgroups | Short for control groups, is a Linux kernel feature that enables fine-grained resource management, allowing processes to be grouped and allocated specific system resources like CPU, memory, and I/O |
| 42  | Microservices patterns | Microservices patterns are architectural guidelines and best practices for designing, deploying, and managing independent, small-scale services that collectively form a larger application |
| 43  | Deployment package | It is a bundled unit containing all necessary files, configurations, and dependencies required to install and run a software application in a specific environment |
| 44  | Dependencies | Dependencies refer to the external libraries, frameworks, or software components required by an application to run within a containerized environment |
| 45  | Exemption cycles | Code foundations have a single API relationship with one another. Concept are that every group may work in isolation and be protected/disconnected from one another |
| 46  | Advanced Multi-Layered Unification File system (AuFS) | Advanced Multi-Layered Unification File system (AuFS) is a union file system used by Docker to layer images and provide a lightweight and efficient way to manage container file systems |
| 47  | rocket container runtime (rkt) | rkt, or Rocket, is an open-source container runtime developed by CoreOS, designed for secure, composable, and standards-focused container deployments |
| 48  | Container orchestration plan | A container orchestration plan outlines how containerized applications will be deployed, managed, scaled, and connected in a cluster, typically using tools like Kubernetes or Docker Swarm |

## 5.1 Comparative Studies

Comparative studies on cloud container systems have scrutinized Docker, LXC, Windows Hyper-V, Podman, runC, and containerd to assess their strengths and limitations. Docker, renowned for its user-friendly interface and extensive ecosystem, excels in rapid deployment and robust image management. LXC, a lightweight alternative, provides low overhead and high performance, but lacks some of Docker's features. Windows Hyper-V, optimized for Windows environments, offers seamless integration with existing Microsoft tools. Podman, a daemonless container engine, emphasizes security and compatibility with Docker. runc and containerd are lower-level runtimes, forming the foundation for Docker and other container platforms, focusing on performance and standard compliance. Each system demonstrates unique capabilities, catering to specific use cases and preferences in cloud-based containerization.

## 6. FINDINGS

The suggested taxonomy, along with extensive technical research and surveys, reveals valuable insights from diverse container systems. These findings offer potential directions for future developments and enhancements in existing systems. Additionally, we address complexities, challenges, and prospects for forthcoming container systems in this context.

### 6.1 *LXC*

Linux Containers (LXC) are a lightweight virtualization technology, allowing multiple isolated Linux systems to run on a single host. Unlike traditional virtual machines, LXC shares the host's kernel, resulting in reduced overhead and enhanced performance. LXC leverages cgroups and namespaces, enabling fine-grained resource control and isolation. It excels in scenarios where efficiency and rapid deployment are paramount, making it popular in development, testing, and production environments. LXC containers boast quick startup times and minimal resource consumption, ideal for microservices architectures and cloud computing. They provide a flexible, efficient solution for orchestrating applications and services, contributing to the versatility and scalability of modern computing environments.

Docker holds several advantages over LXC (Linux Containers) due to its design focus on application-centric containerization. Firstly, Docker's strength lies in its simplicity and user-friendly interface. It abstracts away much of the complexity of containerization, making it accessible to a broader audience. This ease of use has contributed to Docker's widespread adoption and robust community support. Another key advantage is Docker's image management system. It utilizes a layered approach, where containers share common layers, drastically reducing storage requirements and speeding up image pulls. This efficiency in image handling is particularly beneficial for deploying and scaling applications across environments. Docker's ecosystem is a standout feature. It boasts a vast and active community, extensive documentation, and a rich collection of tools and platforms built around it. This ecosystem enhances its versatility and makes it a popular choice for developers and organizations alike.

Portability is also a major strength. Docker containers can run on any system that supports Docker, ensuring a consistent runtime environment regardless of the underlying infrastructure. This portability greatly simplifies application deployment across various environments, from development to production. Moreover, Docker's focus on application packaging is a game-changer. It enables the encapsulation of applications and their dependencies in a standardized format, ensuring that they run consistently across different environments. While LXC has its merits, Docker's emphasis on user-friendliness, efficient image management, a thriving ecosystem, portability, and application-centric containerization makes it a superior choice for many use cases, especially in rapidly evolving and diverse development environments.

### 6.2 *Windows Hyper-V*

Windows Hyper-V is a hypervisor-based virtualization platform by Microsoft. It allows users to create and manage virtual machines on Windows servers. Hyper-V provides robust isolation between virtualized environments, ensuring secure operation of multiple operating systems simultaneously. It supports various guest operating systems, including Windows, Linux, and others. Hyper-V offers features like live migration, enabling seamless movement of virtual machines between hosts without downtime. Its scalability and integration with Windows ecosystems make it a popular choice for enterprises seeking efficient server virtualization solutions. Hyper-V plays a crucial role in consolidating workloads, optimizing resource utilization, and facilitating dynamic IT environments.

Docker and Windows Hyper-V are both powerful virtualization technologies, but they serve different purposes and have distinct strengths. Docker is advantageous for the following reasons. Docker containers share the host OS kernel, resulting in significantly lower overhead compared to full virtualization solutions like Hyper-V. This means Docker containers start faster and consume fewer system resources. Docker containers are highly portable and can run consistently across different environments, crucial for modern DevOps practices and microservices architectures. Docker's containerization model allows for quick deployment and scaling of applications, simplifying the process of packaging an application and its dependencies, ensuring consistent behavior in any environment. Additionally, Docker has a vast and active community, extensive documentation, and a rich ecosystem of tools and platforms built around it, making it easy to find support and resources for Docker-related tasks.

Hyper-V is preferred for certain use cases due to the following reasons. Hyper-V provides complete virtual machines with separate operating systems, essential for running legacy applications or certain specialized workloads. It seamlessly integrates with the Windows environment, making it the natural choice for organizations heavily invested in the Microsoft ecosystem. Hyper-V excels in running Windows-based workloads and is optimized for this purpose. Ultimately, the choice between Docker and Hyper-V depends on specific use cases and the nature of the workloads being managed. Docker excels in containerizing applications for modern, agile development workflows, while Hyper-V provides a full virtualization solution advantageous for Windows-centric environments and legacy applications.

### 6.3 *Podman*

Podman is a containerization tool in the Linux ecosystem, offering a seamless alternative to Docker. What sets Podman apart is its daemonless architecture, eliminating the need for a centralized daemon process. This leads to enhanced security and resource utilization. Podman is compatible with Docker images and containers, simplifying the transition for users familiar with Docker. It provides features like rootless containers, enabling non-privileged users to run containers securely. With support for Kubernetes and Open Container Initiative (OCI) standards, Podman offers a versatile solution for managing containers, making it a valuable tool for developers and system administrators in modern IT environments. Docker has long stood as a cornerstone in containerization technology, boasting several advantages over Podman. Firstly, Docker's widespread adoption and extensive community support have fostered a rich ecosystem of tools and resources, providing users with a wealth of options and expertise. This robust community backing contributes to Docker's stability and reliability.

Additionally, Docker's simplicity and ease of use have made it the go-to-choice for many developers and organizations. Its intuitive command-line interface and straightforward setup process facilitate quick adoption and integration into existing workflows. Docker's layered image system is another significant advantage. By allowing containers to share common layers, it optimizes storage and accelerates image pulls, leading to more efficient resource utilization. Portability is a major strength for Docker. Its containers can run consistently across various environments, providing a uniform runtime environment regardless of the underlying infrastructure. This is crucial for modern DevOps practices and microservices architectures.

Furthermore, Docker's emphasis on application-centric containerization streamlines the process of packaging applications and their dependencies. This ensures consistent behavior in different environments, making it an invaluable tool for development and deployment pipelines. While Podman offers distinct advantages like daemonless architecture and rootless containers, Docker's maturity, extensive ecosystem, user-friendliness, and portability make it a preferred choice for many developers and organizations seeking reliable and efficient containerization solutions.

### 6.4 *runC*

runC, a core component of containerization, is an open-source command-line utility. It adheres to the Open Container Initiative (OCI) specification, enabling the execution of containers on various platforms. RunC is a lightweight, portable tool that spawns and runs containerized applications, acting as a container runtime. Its simplicity and adherence to industry standards make it a crucial building block for container orchestrators like Kubernetes and container management systems like Docker. By providing a standard interface for launching containers, runC enhances compatibility and interoperability in the container ecosystem, allowing developers to create and manage containers without being tied to a specific containerization platform.

Docker and runC serve different but complementary roles in the containerization ecosystem. Docker is a comprehensive container platform that incorporates several components, including runC, to provide a complete containerization solution. However, there are several reasons why Docker, as a complete platform, is often preferred over using runC in isolation. One significant advantage of Docker is its user-friendly interface and high-level abstraction. Docker simplifies the process of creating, managing, and orchestrating containers, making it accessible to a broader audience. It offers a wide range of tools and features that streamline development and deployment workflows, including image management, container networking, and a powerful CLI.

Docker's layered image system is a substantial benefit. It allows containers to share common layers, reducing storage requirements and speeding up image pulls. This efficiency is crucial for deploying and scaling applications in resource-constrained environments. Docker's ecosystem is another major strength. It has a thriving community, extensive documentation, and a rich set of plugins and integrations that enhance its functionality. This ecosystem provides developers and organizations with a wealth of resources and options to enhance their containerized workflows.

Additionally, Docker provides robust support for container orchestration through tools like Docker Swarm and Kubernetes, making it well-suited for managing complex container deployments at scale. While runC is a fundamental component in the containerization landscape, Docker's comprehensive platform, with its user-friendly interface, efficient image management, vibrant ecosystem, and robust orchestration capabilities, often makes it the preferred choice for developers and organizations looking for a complete containerization solution.

### *6.5 Containerd*

Containerd is a core container runtime used in various container platforms. It's designed for stability, performance, and portability, focusing on the essential aspects of container execution. Developed under the guidance of the CNCF, Containerd provides a foundational layer for managing container lifecycle events. It's compatible with industry standards like the Open Container Initiative (OCI) specifications, ensuring seamless integration with other containerization tools. Containerd's modular architecture allows it to be easily extended and customized, making it a versatile choice for orchestrating containerized applications. Its lightweight nature and adherence to best practices make it a popular choice in cloud-native environments.

Docker and Containerd are interconnected components in the containerization ecosystem, with Docker utilizing Containerd as its core container runtime. However, Docker, as a comprehensive platform, offers several advantages over using Containerd in isolation. One key distinction lies in Docker's high-level abstraction and user-friendly interface. Docker simplifies the process of creating, managing, and orchestrating containers, making it accessible to a wider audience. It provides a rich set of tools and features, including image management, container networking, and a powerful CLI, streamlining development and deployment workflows.

Docker's layered image system is a significant asset. By allowing containers to share common layers, it optimizes storage requirements and accelerates image pulls. This efficiency is vital for deploying and scaling applications in resource-constrained environments. The Docker ecosystem is another crucial strength. With a thriving community, extensive documentation, and a wealth of plugins and integrations, Docker offers a wide array of resources and options to enhance containerized workflows.

Moreover, Docker provides robust support for container orchestration through tools like Docker Swarm and Kubernetes. This makes it well-suited for managing complex container deployments at scale. While Containerd serves as a fundamental container runtime, Docker's comprehensive platform, encompassing its user-friendly interface, efficient image management, vibrant ecosystem, and powerful orchestration capabilities, often positions it as the preferred choice for developers and organizations seeking a complete containerization solution.

### 7. CONCLUSION

Container computing is the promising paradigm for delivering IT services as computing utilities. Docker containers are designed to provide services to external users; providers need to be compensated for sharing their resources and capabilities. There are many open issues regarding the Container computing. This proposed taxonomy will provide researcher and developer the ideas on the current container systems, hype and challenges. This paper provides the information to evaluate and improve the existing and new container systems and prefers Docker over others.

### COMPETING INTERESTS

Authors have declared that they have no known competing financial interests OR non-financial interests OR personal relationships that could have appeared to influence the work reported in this paper.

*Peer-review history:*
*The peer review history for this paper can be accessed here:*
*https://www.sdiarticle5.com/review-history/110049*